\documentclass[conference]{IEEEtran}
\IEEEoverridecommandlockouts
\usepackage{cite}
\usepackage{amsmath,amssymb,amsfonts}
\usepackage{verbatim}
\usepackage{algorithmic}
\usepackage[ruled,vlined,linesnumbered]{algorithm2e}
\usepackage{graphicx}
\usepackage{multirow}
\usepackage{textcomp}
\usepackage[dvipsnames]{xcolor}
\usepackage{soul}
\usepackage{boldline}
\usepackage{comment}
\def\BibTeX{{\rm B\kern-.05em{\sc i\kern-.025em b}\kern-.08em
    T\kern-.1667em\lower.7ex\hbox{E}\kern-.125emX}}
    
\makeatletter
\newcommand*\bigcdot{\mathpalette\bigcdot@{.5}}
\newcommand*\bigcdot@[2]{\mathbin{\vcenter{\hbox{\scalebox{#2}{$\m@th#1\bullet$}}}}}
\newcommand{\removelatexerror}{\let\@latex@error\@gobble}
\makeatother
    
\begin{document}

\title{AOBTM: Adaptive Online Biterm Topic Modeling for Version Sensitive Short-texts Analysis\\
\thanks{This research is supported by the Canada NSERC Discovery Grant [RGPIN-2019-05175].}
}

\author{
    \IEEEauthorblockN{Mohammad Abdul Hadi}
    \IEEEauthorblockA{\textit{Department of Computer Science} \\
    \textit{The University of British Comumbia}\\
    Kelowna, Canada\\
    hadi@alumni.ubc.ca}
    \and
    \IEEEauthorblockN{Fatemeh H Fard}
    \IEEEauthorblockA{\textit{Department of Computer Science} \\
    \textit{The University of British Comumbia}\\
    Kelowna, Canada\\
    fatemeh.fard@ubc.ca}
}

\maketitle
\begin{abstract}
    Analysis of mobile app reviews has shown its important role in requirement engineering, software maintenance and evolution of mobile apps. 
    Mobile app developers check their users’ reviews frequently to clarify the issues experienced by users or capture the new issues that are introduced due to a recent app update. 
    App reviews have a dynamic nature and their discussed topics change over time. 
    The changes in the topics among collected reviews for different versions of an app can reveal important issues about the app update. A main technique in this analysis is using topic modeling algorithms.
    However, app reviews are short texts and it is challenging to unveil their latent topics over time. 
    Conventional topic models such as Latent Dirichlet Allocation (LDA) and Probabilistic Latent Semantic Analysis (PLSA) suffer from the sparsity of word co-occurrence patterns while inferring topics for short texts.
    Furthermore, these algorithms cannot capture topics over numerous consecutive time-slices (or versions). Online topic modeling algorithms such as
    Online LDA (OLDA) and Online Biterm Topic Model (OBTM) speed up the inference of topic models for the texts collected in the latest time-slice by saving a fraction of data from the previous time-slice. But these algorithms do not analyze the statistical-data 
    (such as topic distributions) 
    of all the previous time-slices, which can confer contributions to the topic distribution of the current time-slice. 
    
    In this paper, we propose Adaptive Online Biterm Topic Model (AOBTM) to model topics in short texts adaptively. 
    AOBTM alleviates the sparsity problem in short-texts and considers the statistical-data for an optimal number of previous time-slices. 
    We also propose parallel algorithms to automatically determine the optimal number of topics and the best number of previous versions that should be considered in topic inference phase. 
    Automatic evaluation on collections of app reviews and real-world short text datasets confirm that AOBTM can find more coherent topics and outperforms the state-of-the-art baselines. For reproducibility of the results, we open source all scripts.

\end{abstract}

\begin{IEEEkeywords}
App review analysis, adaptive topic model, biterm, online algorithm, automatic parameter setting
    \vspace{-3mm}
\end{IEEEkeywords}

\section{Introduction}
\label{sec;intro}
    \vspace{-1mm}
    Mobile app reviews form a main feedback channel for the app developers \cite{gao2018infar} to evaluate their products and improve application maintenance and evolution tasks \cite{SURMINER}. The app developers require to analyze app reviews in order to gain insights about the current state of their apps from users' perspectives. 
    Mobile app reviews form a feedback channel for the developers \cite{gao2018infar} to evaluate their products and improve application maintenance and evolution tasks \cite{SURMINER}. The app developers must analyze app reviews in order to gain insights about the current state of their apps from users' perspectives. 
    Several studies have been proposed to analyze the app reviews, including extracting informative text \cite{chen2014ar}, summarizing user reviews \cite{martin2016survey}, identifying the bugs or feature requests \cite{palomba2015user, pagano2013user, maalej2015bug}, prioritizing feature inclusions \cite{devReq}, and extracting insights about the apps \cite{gao2018infar}. 
    Although the (popular) apps are updated frequently \cite{diver}, the app review analysis studies mostly consider the app reviews static \cite{idea, hassan2018studying}. 
    However, app reviews have a dynamic nature and their discussed topics change over time.
    If the update-centric analysis is neglected, it misses the point that feedback are written on a certain update \cite{hassan2018studying}. 
    The change in the topics extracted from reviews for different app versions can reveal important issues about the app \cite{idea, hassan2018studying}. 

    A recent example of the importance of discussed topics over time is the \textit{Zoom Cloud Meeting}, a popular app for video conferencing. Zoom received massive one-star ratings (the lowest rating) in Google Play Store during the COVID-19 outbreak in March 2020. Most of the issues were related to users' concerns about data-privacy and security-malpractices. These issues were so severe that in a letter to Zoom, the New York attorney general’s office expressed concerns and addressed security flaws \cite{zoom}. These issues were extensively discussed in the user reviews and could have been flushed out months ago through proper inspection of the user reviews and topic changes from user feedbacks on Google Play Store. 

    App reviews are short texts that are time/version sensitive as these texts are generated constantly and are collected regularly for consecutive app versions \cite{idea, diver}. The underlying latent topics derived from app reviews can benefit the developers extensively \cite{noei2019too}. 
    However, extracting relevant topics from app reviews is challenging due to their dynamic nature and lack of rich context in short texts. 

    The most popular topic modeling methods for discovering the underlying topics from text-corpus are LDA \cite{lda} and PLSA \cite{plsa}. But these topic models do not perform well with text-corpus containing short-texts as documents \cite{btm}. 
    These algorithms consider individual short texts as separate documents and model each of these documents as a mixture of topics, where each topic is considered as a probability distribution over words. The models then utilize various statistical techniques to determine the topic components and mixture coefficients of each document by implicitly capturing the document-level word co-occurrence patterns \cite{es1, tot}. 
    While dealing with the typical lengthy documents, the mentioned algorithms could rely on larger word counts to know how the words are related. However, the natural sparseness of the word co-occurrence patterns in each short document makes these models suffer from the data-sparsity problem \cite{sparse}. Moreover, short texts lack the richness of context, making it more difficult for these topic models to identify the senses of ambiguous words in short documents. 
    Biterm Topic Model (BTM) alleviates these problems by learning topics over short texts and explicitly modeling the generation of biterms in the whole corpus to enhance topic learning \cite{btm}.
    
    As mentioned, app reviews are usually collected in batches of consecutive time-slices \cite{idea}. Each time a new batch of text-data arrives, these topic models (e.g. LDA, BTM) require retraining to discover latent topic distributions from the new dataset, which is prohibitively time and memory consuming. The popular way to alleviate the scalability problem is to develop online algorithms such as Online LDA (OLDA) \cite{olda} and Online BTM (OBTM) \cite{btm}. These online algorithms store a small fraction of data on the fly in order to accommodate the dataset of the upcoming time-slice. When a new batch of text-data arrives, online algorithms model the topics of texts either by using the statistics of samples collected in the immediately previous version/time-slice\footnote{Time-slice and version are semantically equal in this paper.} \cite{obtm} or by naively aggregating statistics of all the previous time-slices \cite{btm}. However, these online algorithms do not take different versions' varying contributions into account. Statistics of the textual data collected over different time periods or different versions may have a non-negligible difference in similarity with that of the latest time-slice; and thus, can contribute differently to the latest version\cite{tot, idea}. Adaptive versions of online algorithms, such as Adaptively Online LDA (AOLDA), can be used to address the problem of the varying contribution of different versions \cite{idea}. But, the underlying model in AOLDA is LDA, which again makes it suffer from the mentioned data-sparsity problem while working with short-texts.
    
    In this paper, we propose a new adaptive online topic model for short texts which takes previous versions' varying contribution into account. We refer to this novel model as the \textbf{A}daptive \textbf{O}nline \textbf{B}iterm \textbf{T}opic \textbf{M}odel (AOBTM). AOBTM inherits the characteristics of BTM to deal with the data sparsity issue. It is an online algorithm that can scale for the increasing volume of the dataset that is generated frequently. AOBTM also endows the statistics of the previous versions with different contributions to the topic distributions of the current version of the dataset. Also, we have employed a preprocessing technique that is useful for yielding better top contributing key-terms to help the manual investigation of the inferred topics.
    Our contributions are enlisted below:
    \begin{enumerate}
        \item We propose a novel method called AOBTM for version sensitive content analysis for short texts. This method adaptively combines the topic distributions of a selected number prior versions to generate topic distributions of the current version.
        \item We propose two parallel algorithms; the first algorithm can identify an optimal number of topics to be derived in the latest version, and the second algorithm can identify the optimal number of previous versions to be taken into consideration for adaptive aggregation of statistical data.
        \item To  encourage  replicability, 
        of the research results, 
        we make all scripts, codes, and graphs available to the community\footnote{https://anonymous.4open.science/r/995c2443-74d9-4e10-a3fa-4f814082b06d/}.
    \end{enumerate}

    We have conducted experiments on app review datasets and Twitter dataset with large number of records to evaluate performance of AOBTM compared to five baseline algorithms. 
    Also, we integrated AOBTM into the state of the art online app-review analysis framework called IDEA for comparison \cite{idea}.
    Our 
    results show that topics captured by AOBTM are more coherent compared to the topics extracted by baseline methods.

    The rest of the paper is organized as follows: Section \ref{sec:background} and \ref{sec:aobtm} describe the background and our Topic Model Design. Sections \ref{sec:optTopNum} and \ref{sec:experiments} are dedicated to the proposed parallel algorithms and experiments and results, followed by the related works in Section \ref{sec:relatedWork}. We add threats to validity in section \ref{sec:threats} and conclude the paper in Section \ref{sec:conclusion}.
    \vspace{-1mm}

\section{Background and Motivation}
\label{sec:background}
\subsection{Topic Modeling for conventional text-documents}
\vspace{-1mm}
    Topic modeling algorithms such as PLSA and LDA are widely embraced for identifying latent semantic structures from text corpus without requiring any prior annotations of the documents \cite{survey}. These algorithms observe each document as a mixture of topics while a distribution over the vocabulary terms characterizes each topic. 
    Statistical techniques such as Variational methods and Gibbs sampling are then applied to infer the latent topic distributions of given documents and the word distributions of inferred topics \cite{ptm}. 
    Although these algorithms and their variants contributed largely in modeling text collections such as blogs, research papers, and news articles \cite{olda, idc, iwck}, these topic models endure considerable performance deterioration while handling short texts \cite{dstm, btm}.
    Directly applying these models on short texts suffer from severe data sparsity problem \cite{sparse} as the frequency of words in the short texts play less discriminative role, which makes it hard to infer words correlation from short documents \cite{sparse}. The limited contexts in the short text also make it challenging to identify the sense of ambiguous words. 

\vspace{-2mm}
\subsection{Topic Modeling for short text-documents}
\vspace{-1mm}
    Researchers have proposed the numerous topic modeling algorithms for short texts by trying to solve one or two of the following inherent characteristics of the short texts: 
\textit{i) lack of enough word co-occurrence information, probability of most individual short texts being generated by singular topic, 
ii) inability to fully capture semantically co-related but rarely co-occurring words (due to lack of statistical information of words among texts), and 
iii) the probability that a single-topic assumption is too strong for some short texts} \cite{survey}.
    In \cite{survey}, Qiang et al. divided the short text topic modeling algorithms into three major categories: 
\textit{Dirichlet multinomial mixture (DMM) based methods,
Global word co-occurrences based methods, and 
Self-aggregation based methods}. 
Brief introductions to the related works can be found in section \ref{sec:relatedWork}.

\vspace{-2mm}
\subsection{Online Topic Modeling for short texts}
\vspace{-1mm}
    A particular issue with the traditional topic modeling algorithms is that they can not scale with the expanding dataset. Whenever a new batch of data arrives, these topic models (i.e., LDA, PLSA) need to train from scratch. Moreover, these conventional topic models can not guarantee consistency in the sequence of topics if independent training is performed on different batches of the same corpus \cite{lda, btm}. This inconsistency occurs because, before each of the independent training, we set the prior topic distribution to a default value, a Dirichlet parameter; so, the fixed number of topics (assume, $K$) can be generated in any sequence. \cite{btm}. 
    For example, when we train a corpus, using BTM or LDA, with $K$ number of topics, each independent training of BTM generates $K$ number of topic distributions, where we can not ascertain that each time we train the corpus, a topic no. $K=k$ (here, $k=[1,...,K]$) will always correspond to a specific topic (i.e., "UI\_component").

    Researchers proposed online models (i.e., OBTM, OLDA) to circumvent the problems with streaming datasets. Here, we can assume that the documents would be generated in streams and can be collected from, divided different time-slices or versions, where the documents are exchangeable in a time-slice.    
    For example, Online BTM (OBTM) accommodates and deals with batches of short-text documents divided into different time-slices or versions. Let's assume that OBTM has already got the topic distribution for $(t-1)$-th time-slice. When a new batch ($t$-th time-slice) arrives, OBTM utilizes the topic distribution of $(t-1)$-th time-slice to set the prior topic distribution of $t$-th time-slice. It, in turn, ensures that after the training of the $t$-th time-slice, the $k$-th topic in the $t$-th time-slice is closely related to the $k$-th topic generated in the $(t-1)$-th time-slice. If we introduce a completely new topic in the latest batch of documents ($t$-th time-slice), the new topic will merge into one (or more) existing topic(s) generated in the $(t-1)$-th time-slice that has (/have) strong correlation(s) with the introduced topic. \cite{btm}
    
\vspace{-2mm}
\subsection{Adaptively Online Topic Modeling for short texts}
\vspace{-1mm}
    The problem with online topic modeling algorithms is that they do not consider or compare the varying consequential correlation among all the preceding time slices or versions of the short texts while inferring topics for the latest time-slice. For example, in OBTM, this limitation transpires as the topic model generates the topic distribution of a time-slice by making it directly dependent on the topic distribution of preceding time-slices. If new topics are being introduced in each time-slice, the $k$-th topic in the latest time slice would be significantly different than that of the first time-slice. We can not reliably compare the distribution of the $K=k$-th topic between two non-consecutive time-slices as OBTM does not impose or investigate the varying correlation between them.
    
	In our proposed method, we aim to alleviate the mentioned problem by \textit{adaptively} integrating the topic distributions of all the previous time-slices with their respective weights as contributions, for generating the prior distribution of the latest, $t$-th time-slice. This way, we can warrant both coherence of specific topics and consistency in the topics' sequence in all the available versions. The adaptiveness enables us to compare topic distributions of any two different time-slices reliably.
    \vspace{-2mm}
    
    \begin{figure}[hbtp]
        \centering
        \includegraphics[width=0.7\linewidth]{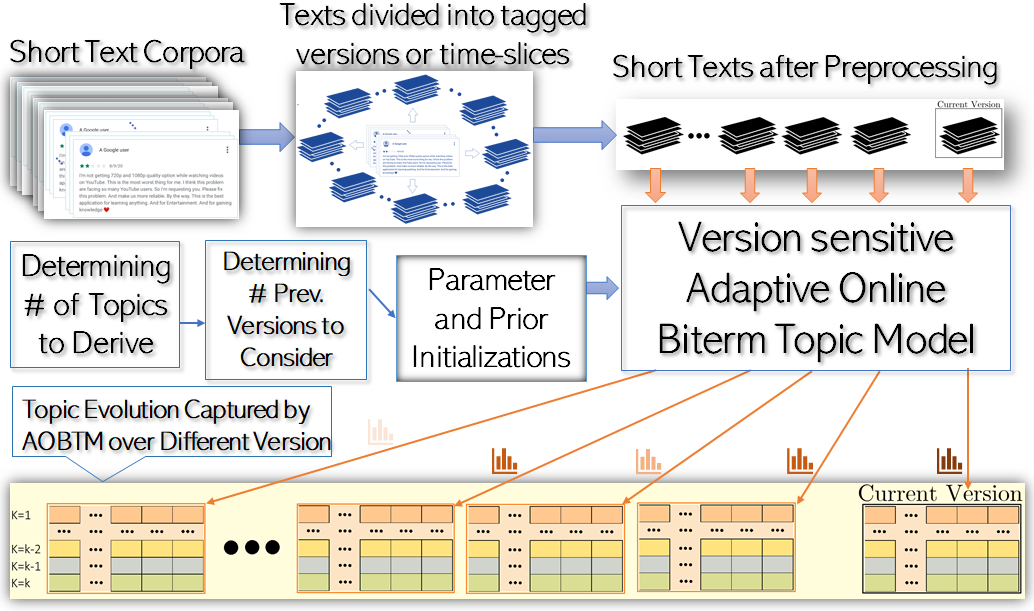}
        \caption{Overview of the Framework}
        \label{fig:overview}
        \vspace{-3mm}
    \end{figure}

    In Figure \ref{fig:overview}, we show an overview of the framework. Here, version tagged short-texts are processed and fed into the AOBTM algorithm to find better topic distribution for the latest version by leveraging previous versions' statistical data. The details of each part are discussed in Sections \ref{sec:aobtm}, \ref{sec:optTopNum}.
    
\vspace{-1mm}
\section{Adaptive Online Biterm Topic Model}
\vspace{-1mm}
\label{sec:aobtm}
    In this section, we discuss the details of the Adaptive Online Biterm Topic Model (AOBTM). 
    This method introduces \textit{Adaptiveness} to give the online algorithm, OBTM, a version or time-slice sensitivity so that the prior topic distribution of the latest time-slice takes varying contributions (topic distribution-wise) of the previous time-slices into account. After setting the prior, we train the model to find out the final topic distribution. The details of the proposed method are described below.
\vspace{-2mm}
\subsection{Applied Biterm Extraction technique}
\vspace{-1mm}
    We have adopted the definition of \textit{Biterm} from \cite{btm}, where it denotes an unordered term-pair co-occurring in a small, fixed-size window over a term sequence. The fixed-sized window is referred to as \textit{short-context}. The optimal size of short-context varies from dataset to dataset and can be considered as an important parameter setting. In a given short-context, an unordered pair of any two distinct terms can form a biterm. For example, a short context with size=3, generates the following biterms:
    	$(w_1,w_2,w_3) \Rightarrow \{(w_1,w_2), (w_2,w_3), (w_1,w_3)\}$

    In previous works of topic labeling (i.e., \cite{tl1, tl2}), inferred topics have been labeled with the term(s), which has(/have) a more significant contribution to the respective topics. Here, \textit{term} denotes any non-redundant word in the document, which cannot be found in Natural Language Toolkit's (NLTK) stop-word list. But Gao et al. \cite{diver}, showed that the most contributing singular term or their combination could not adequately represent the respective topic. So, instead of using singular terms, we use meaningful phrases to label the topics, where a \textit{Phrase} refers to two frequently co-occurring words. To ensure the comprehensibility of the extracted phrases, we use a Pointwise Mutual Information (PMI)- based phrase extraction method \cite{pmi}, where the higher frequency of two words' co-occurrence warrants the generation of a more meaningful phrase. 
    For our model, we have empirically set our frequency threshold to 24.
    After identifying the phrases, we convert them into single terms using `\_' (i.e., $w_1\_w_2$), to train them along with other terms using our algorithm. During biterms extraction, words constructing the phrases are also considered as \textit{term} when they appear outside identified phrases. We train the phrases to capture their underlying semantics, which, in turn, would help us to label the topics with the most relevant phrases. We will further demonstrate this modification's impact in the experiment section. 
\vspace{-1mm}
\subsection{Model Description}
\begin{figure}[hbtp]
        \centering
        \includegraphics[width=0.85\linewidth]{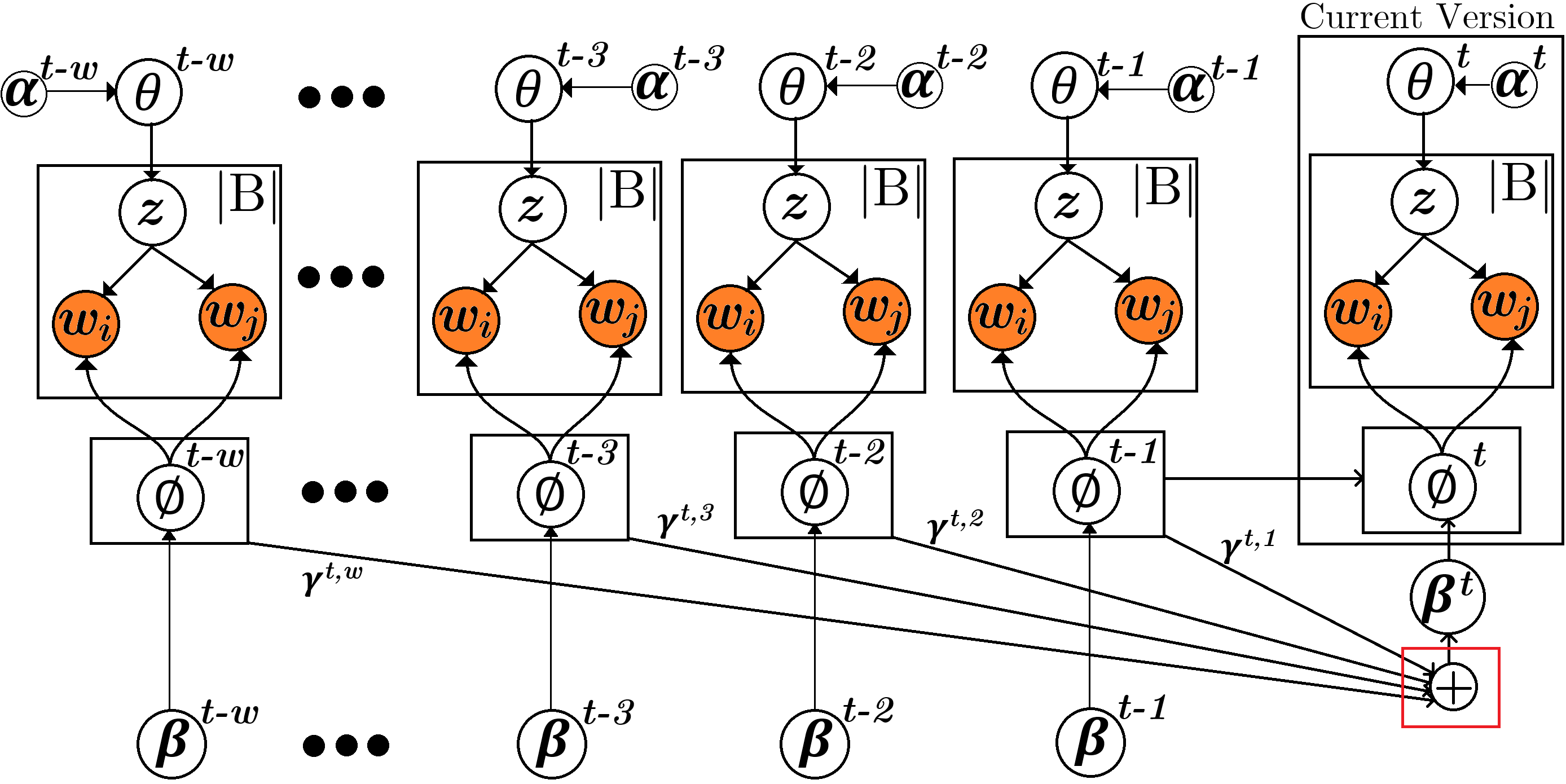}
        \caption{Overview of AOBTM. The red rectangle highlights the adaptive integration of the topics of the \textit{w} previous versions for generating the prior $\Phi$ in the t-th version}
        \label{fig:aobtm}
        \vspace{-5mm}
    \end{figure}

\vspace{-1mm}
    To alleviate the data-sparsity problem faced by AOLDA and to capture more coherent, comprehensible, and discriminative topics, we propose an adaptive online topic modeling method, AOBTM, which improves OBTM by adaptively combining the topic distributions in previous versions. 
    The details of the proposed AOBTM method are described in figure \ref{fig:aobtm}.
    
    After the preprocessing, the short texts are separated into different time-slices or versions, and input into AOBTM sequentially. AOBTM treats the short texts-set from each time-slice as a separate corpus. 
    We denote the whole corpus as $R$ = {$R^1$, $R^2$, ..., $R^t$}, where t indicates the t-th time-slice.  Following the literature, we denote the prior distributions over corpus-topic as $\alpha$ and the prior distributions over topic-words as $\beta$; both $\alpha$ and $\beta$ are defined initially.  The topic-word distributions determine the topic's distribution over all the non-redundant terms (including the phrases) that appear in the corpus.
    The number of the topics is specified as K. For the k-th topic, $\Phi_k^t$ is the probability distribution vector over all the input terms in the $t$-th time slice. We introduce a new parameter- $win$ (window size), which defines the number of previous versions to be considered for inferring the topic distributions of the current version. The overview of the AOBTM model is depicted in Figure \ref{fig:aobtm}.
    Different from OBTM (and similar to AOLDA), as Figure \ref{fig:aobtm} shown, we adaptively integrate the topic distributions of the previous $win$ versions, denoted as {$\Phi^{t-1}, \Phi^{t-2}, ..., \Phi^{t-i}, ..., \Phi^{t-win}$}, for generating the prior, $\beta^t$ for the t-th version. The adaptive integration sums up the topic distributions of different versions with different weights, $\gamma^{t,i}$:
\vspace{-3mm}
    \begin{equation}
    \label{eq:betaUp}
\resizebox{0.5\hsize}{!}{$
        \beta_k^t = 
                    \sum_{i=1}^{win} \gamma_k^{t,i} \Phi_k^{t-i} 
                    + n_{w|k}^t
    $}
    \vspace{-1mm}
    \end{equation}
    Here, $i$ denotes the $i$-th previous version ($1 \leq i \leq w$). $n_{w|k}^t$ denotes the number of times word, w is assigned to topic k in time-slice t. The weight $\gamma_k^{t,i}$ is determined by considering the similarity of the k-th topic between the (t-i)th version and the (t-1)th version, which is calculated by the following softmax function:
\vspace{-3mm}
    \begin{equation}
    \label{eq:soft}
\resizebox{0.5\hsize}{!}{$
        \gamma_k^{t,i} = \frac{exp(\Phi_k^{t-i} \cdot \beta_k^{t-1})} {\sum_{j=1}^{win} exp(\Phi_k^{t-j} \cdot \beta_k^{t-1})}
    $}
    \end{equation}

    \scalebox{0.95}{
    \removelatexerror
    \begin{algorithm}[H]
    \caption{Adaptive Online BTM}
    \label{alg:aobtm}
    \SetAlgoLined
    \SetKwInOut{Input}{Input}
    \SetKwInOut{Output}{Output}
    \SetKw{KwBy}{by}
    \SetKw{KwTo}{to}
    \Input{$K, win, \alpha, \beta, \mathbf{B^{(1)},...., \mathbf{B^{(T)}} }$}
    \Output{\{$\Phi^{(t)}, \theta^{(t)}\}_{t=1}^{T}$}
    Set $\alpha^{(1)} = (\alpha,....,\alpha)$ and $\{\beta_{k}^{(1)} = (\beta,....,\beta)\}_{k=1} ^{K}$\\
    \For{$t \gets 1$ \KwTo $T$}{%
        Randomly assign topics to biterms in $B^{(t)}$\;
        \For{$iter \gets 1$ \KwTo $N_{iter}$}{%
            \ForEach{biterm $b_i = (w_{i,1},w_{i,2}) \in B^{(t)}$}{%
                    Draw topic k from \textbf{Eq. \ref{big}}\;
                    Update $n_{k}^{(t)}, {n_{w_{i,1}}}_{|k}^{(t)}$, and ${n_{w_{i,2}}}_{|k}^{(t)}$\;
            }
            Set $\alpha^{(t+1)}$ \KwBy \textbf{Eq.\ref{eq:alpha}}\;
            Set $\{\beta_{k}^{(t+1)} \}_{k=1}^{K}$ \KwBy \textbf{Eq.\ref{eq:betaUp}}\;
        }
        Compute $\Phi^{(t)}$ \KwBy $\phi_{k,w} = \frac{n_{w|k} + \beta,}{n_{\cdot|k} + \boldsymbol W \beta}$; (refer to \cite{btm})\\
        Compute $\theta^{(t)}$ \KwBy $\theta_k = \frac{n_k + \alpha,}{N_B + \boldsymbol K \alpha}$; (refer to \cite{btm})
    }
    \end{algorithm}
    }

	In Equation \ref{eq:soft}, [$\Phi_k^{t-i} \cdot \beta_k^{t-1}$] represents \textit{Einstein Summation} and computes the similarity between the topic distribution, $\Phi_k^{t-i}$ and the prior of the (t-1)th version, $\beta_k^{t-i}$. This adaptive aggregation allows the topics of the previous versions to endow different contributions to the topic distributions of the current version. The steps are shown in Algorithm \ref{alg:aobtm}. In Algorithm \ref{alg:aobtm}, $\mathbf{B}$ denotes the biterms collection. Here, $N_B$ is the number of biterms and $b_i$ denotes a biterm with two terms: $w_{i,1}$ and $w_{i,2}$ in $i$-th biterm. 
	We use $\mathbf{W}$ as the total number of words in the vocabulary, and
	$\mathrm{\mathbf{\theta}^{(t)}}$ as a K-dimensional multinomial distribution which denotes the corpus-topic distribution for a time-slice. Here, $n_k$, $n_{k|d}$, and $n_{w|k}$ denote number of words in topic k, number of words in document d assigned to topic k, and number of times word w is assigned to topic k, respectively.
	
    In Algorithm \ref{alg:aobtm}, the topics are drawn from Eq. \ref{big} and  prior distribution, $\alpha$ for the latest time-slice is calculated using Eq.\ref{eq:alpha}:
    \begin{align}
        P& \big(z_{i} = k|z_{-i}^{(t)}, B^{(t)}, \alpha^{(t)}, \{\beta _{k}^{(t)} \}_{k=1}^{K}\big) \nonumber \\
        &\propto (n_{-i,k}^{(t)} + \alpha _{k}^{(t)}) \frac{(n_{-i,w_{i}|k} + \beta_{k, w_{i}}^{(t)}) (n_{-i,w_{j}|k} + \beta_{k, w_{j}}^{(t)})}{[\sum_{w=1}^{W} (n_{-i,w|k} + \beta_{k, w}^{(t)})]^{2}} 
    \label{big}
    \end{align}
\vspace{-4mm}
    \begin{equation}
    \label{eq:alpha}
\resizebox{0.3\hsize}{!}{$
        \alpha_{k}^{(t+1)} = \alpha_{k}^{(t)} +n_{k}^{(t)}
    $}
        \vspace{-1mm}
    \end{equation}
    where, $z \in [1,K]$ refers to the topic indicator variable and P(z) refers to the prevalence of topics in the corpus. We use symmetric Dirichlet distributions as the initial priors by setting $\alpha^1 = (\alpha, ..., \alpha)$ and $\beta_k^1 = (\beta, ... , \beta)$. Given $\alpha^t$ and $[{\beta_k^t}]_{k=1}^K$, we iteratively draw topic assignments for each biterm $b_i \in \mathbf{B}^t$, according to the conditional distribution stated in Eq. \ref{big}. Once iterations are completed, we obtain the counts $n_k^t$ and $n_{w|k}^t$. We adjust the hyperparameters $\alpha^t$ and $[{\beta_k^t}]_{k=1}^K$ for time slice $(t+1)$ by setting $\alpha_{k}^{(t+1)}$ and $\beta_k^{t+1}$ using Eq. \ref{eq:alpha} and Eq. \ref{eq:betaUp}, respectively. The derivation of Eq. \ref{big} and Eq. \ref{eq:alpha} can be found in \cite{btm}.
    
\vspace{-2mm}
\subsection{AOBTM Complexity and Comparison with Baselines}
\vspace{-1mm}
    In this section, we discuss the details of the running time and memory requirement for AOBTM and compare it with different batch, online, and adaptive online algorithms. We have listed the time-complexity and the number of in-memory variables for different topic models in Table \ref{table:complex}. 
    
    In the following discussion, $\Bar{l}$ refers to the average document length, and $N_D$ refers to the number of documents in the corpus, respectively. We can assume that all the documents in the short-text corpus have almost the same length \cite{btm, obtm}. It is reasonable to infer $N_B$ (number of biterms in the corpus) using this assumption as we are applying $N_B$ only for the topic models, which are devised for short texts (i.e., BTM, OBTM, and AOBTM). According to our assumption, each document with length $\Bar{l}$, would produce $\Bar{l}(\Bar{l}-1)/2$ biterms; so, we have the equivalence of $N_B$ as:
        $ \approx N_D \cdot \Bar{l} \cdot (\Bar{l}-1) / 2$
    
    Furthermore, in Table \ref{table:complex}, $win$ denotes the user-defined \textit{window-size} (the number of previous versions to consider) in the adaptive inline algorithms, $W$ denotes the total number of terms, and $v$ refers to the number of available time-slices.
    \begin{table}[htbp]
    \caption{Time Complexities and the Number of In-Memory Variables in Different Topic Models}
    \begin{center}
    \renewcommand{\arraystretch}{1.5}
    \resizebox{0.8\linewidth}{!}{%
    \begin{tabular}{l|l|l}
        \hline
        Methods & Time Complexities & \# of Variables in Memory\\
        \hline
        LDA & $O(N_{iter} K N_D \Bar{l})$ & $N_DK + WK + N_D\Bar{l}$\\ 
        BTM & $O(N_{iter} K N_B)$ & $K + WK + N_B$\\
        OLDA & $O(N_{iter} K |N_D^{(t)} \Bar{l}^{(t)}|)$ & $N_DK+WK+|N_D^{(t)} \Bar{l}^{(t)}|$\\  
        OBTM & $O(N_{iter} K | N_B^{(t)}|)$ & $K+WK+|N_B^{(t)}|$\\
        AOLDA & $O(N_{iter} K |N_D^{(t)} \Bar{l}^{(t)}|+vKW)$ & $N_DK+vWK+|N_D^{(t)} \Bar{l}^{(t)}|$\\
        AOBTM & $O(N_{iter} K | N_B^{(t)}|+vKW)$ & $K+vWK+|N_B^{(t)}|$\\
        \hline
    \end{tabular}}
    \end{center}
    \label{table:complex}
\vspace{-5mm}
    \end{table}
    
    \textit{Time Complexity.}
    The most time-consuming part in these topic models is the component calculating the conditional probability of topic assignments, which requires $O(K)$ time. While LDA draws a topic for each word occurrence, BTM draws a topic for each biterm. So, the overall time-complexity for LDA and BTM turn out as $O(N_{iter} K N_D \Bar{l})$ and $O(N_{iter} K N_B)$, respectively \cite{lda, btm}. From our previous assumption-based calculation of $N_B$, we can further expand the time-complexity for BTM: $O(N_{iter} K N_D \Bar{l} (\Bar{l} -1)/2)$, which is approximately $(\Bar{l} -1)/2$ times the time-complexity of LDA. As BTM works with short texts where value of $\Bar{l}$ is considerably small, the run-time of BTM can still be compared to that of LDA \cite{obtm}.
    
    The online algorithms, such as OLDA and OBTM, deal with documents and short texts, respectively, present in the latest time-slice. In Table \ref{table:complex}, we have used superscript $t$ to denote the latest time-slice or version. But, the adaptively online algorithms (i.e., AOLDA, AOBTM) compare and determine contributions of the previous $v$ number of topic-word distributions for different time-slices, which require an additional $O(vKW)$ time.
    
    \textit{Number of Variables Stored in Memory.}
    LDA maintains the following counts as the cached memory: the number of words in a document $d$ assigned to topic $k$, $n_{k|d}$ (=$N_DK$), and the number of times word $w$ assigned to topic $k$, $n_{w|k}$ (=$WK$). LDA also stores the topic assignment for each word occurrence (=$N_D\Bar{l}$) \cite{ldacm}.
    On the other hand, BTM stores the following variables: the number of topics, $n_k$ (=$K$), the number of times word $w$ assigned to topic $k$, $n_{w|k}$ (=$WK$), and the topic assignment for each biterm (=$N_B$) \cite{btm}.
    
    Unlike the batch algorithms, online topic models do not require running over all documents (in case of OLDA), or all biterms (in case of OBTM) observed up to the latest time slice. Instead, OLDA only iteratively runs over the words present in the current time-slice documents, whereas OBTM only iterates over the biterm set in the latest time-slice. These online algorithms require almost constant memory cost to update the models, since the number of documents, their average length, and the number of biterms are often stable \cite{obtm, olda}.
    
    In the adaptively online algorithms, topic-word assignments for different versions are compared, weighted, and combined to set the prior topic-word distribution of the latest time-slice. Therefore, the counts, $n_{w|k}$ (=$WK$) for all the previous time-slices, need to be stored as cache-memory. As $win \in [1,...,v]$, we consider $vWK$ as the counts stored in memory.
    
    From table \ref{table:complex}, we can see that 
    AOBTM's time complexity is higher, but it is comparable to other algorithms while dealing with a fewer number of short texts. In practice, the number of texts is bound to decline as they are separated into different versions or time-slices. 
    On the other hand, AOBTM has to store some additional variables to accommodate \textit{adaptiveness}, yet incur less memory cost than the other Adaptive Online algorithm, AOLDA.

\section{Algorithms to Find Optimal Number of Topics to Infer and Previous Versions to consider}
\label{sec:optTopNum}

    Two parameters in the adaptive online topic modeling method, play key-roles in the quality of the topics discovered: (i) the number of topics to derive, (ii) and the number of previous versions to consider for adaptive integration. In previous studies, the values of these crucial parameters were set via informed guess established from the manual examinations performed over the dataset \cite{idea}. We propose algorithms to determine the values of these parameters automatically. Before developing algorithms to find optimal values for these parameters, 
    we need to determine suitable evaluation metric for measuring the quality of discovered topics. 
    
    Perplexity (or, marginal likelihood) 
    evaluated on a held-out test set have been utilized in many studies to assess the effectiveness and efficiency of a generated topic model \cite{lda, htmm, its}. But, the \textit{minimized perplexity} as a metric is not suitable for our approach for the following reasons. First, the mentioned studies focused on LDA-based topic models where the likelihood of word occurrences in documents is optimized, whereas, in our approach, the likelihood of biterm occurrences in the latest time-slice is optimized. Second, it was argued in \cite{tea}, that topic models with better held-out likelihood might infer less semantically meaningful topics, which deviates our underlying expectations of topic models (e.g., better interpretability and coherence of the derived topics).
    
    For our purpose, we can use \textit{Coherence Score} or \textit{PMI-Score}. \textit{Coherence Score} is a metric used for measuring the quality of the discovered topics automatically \cite{coherence}. It depicts that a topic is more coherent if the most probable words in that topic co-occur more frequently in the corpus. On the other hand, \textit{PMI-Score} measures the coherence of a topic based on point-wise mutual information using large scale text datasets from external sources, such as Wikipedia \cite{pmi}. This idea resonates with the underlying assumption of our approach, which maintains that words co-occurring more frequently in an external dataset, should be more likely to belong to the same topic. Since the external dataset is model-independent, the generated PMI-Score would fluctuate consistently for distinct topic models with different parameter values \cite{btm}. Therefore, we exploit PMI-Score to evaluate the discovered topic quality, which measures the pairwise association among $T$ most contributing words in a discovered topic, $k$:
\vspace{-1.5mm}
    \begin{equation}
    \label{eq:pmi}
\resizebox{0.8\hsize}{!}{$
        \mbox{PMI-Score}(k) = \frac{1}{T(T-1)}\sum_{1 \leq i < j \leq T} log \frac{P(w_i, w_j)}{P(w_i) P(w_i)}
    $}
\vspace{-0.5mm}
    \end{equation}
    Here, $P(w_i)$, $P(w_j)$, and $P(w_i, w_j)$ are the probabilities of word $w_i$, $w_j$, and co-occurring word-pair $(w_i, w_j)$, respectively. The probabilities are estimated empirically from a fixed external dataset. Following the literature \cite{btm, obtm}, we computed the PMI-Scores using 5.4 million English Wikipedia articles as external dataset. We have used an open source web-scraper API \footnote{https://github.com/martin-majlis/Wikipedia-API} to scrape the articles with average length of 362.7 words. To determine the overall PMI-Score for the topic model, we take the average of all PMI-Scores produced by distinct topics: PMI-Score(terminal) $= \frac{1}{K}\sum_k \mbox{PMI-Score}(k)$.
\vspace{-3mm}
\subsection{Algorithm to Determine Appropriate Topic Number}
\vspace{-1mm}
    An inadequate number of topics could render our topic model too coarse to identify distinct and particular topics. Conversely, an inordinate number of topics could deliver a model that is too involved, making subjective validation and interpretation difficult.
    
    \scalebox{0.8}{
    \removelatexerror
    \begin{algorithm}[H]
    \caption{Optimal Number of Topics}
    \label{alg:optTopNum}
    \SetAlgoLined
    \SetKwInOut{Input}{Input}
    \SetKwInOut{Output}{Output}
    \SetKw{KwBy}{by}
    \SetKw{KwTo}{to}
    \Input{$InputArr, iter, span, dataset$}
    \Output{$optTopicNum$}
    Set $maxPMI \gets 0.0$, $optVal \gets InputArr[0];$\\
    \textit{\#pragma omp parallel for reduction(max:maxPMI})\\
    \For{$i \gets 0$ \KwTo $InputArr.size()-1$}{%
        //Each thread works on 1 element of $InputArr$\\
        PMI\_sum $\gets$ 0.0;\\
        \For{$j \gets 0$ \KwTo $iter-1$}{
            Set $threadId \gets omp\_get\_thread\_num();$\\
            Set $topicNum \gets InputArr[threadId];$\\
            Set $K$ in Algorithm \ref{alg:aobtm} with $topicNum;$\\
            \{$\Phi^{(t)}, \theta^{(t)}\}_{t=1}^{T} \gets$ Run AOBTM (Algorithm \ref{alg:aobtm});\\
            PMI[$K$] $\gets$ new array of double;\\
            \For{$t \gets 0$ \KwTo $K-1$}{%
                PMI[$t$] $\gets$ \KwBy Eq. \ref{eq:pmi};\\
            }
            PMI\_score $\gets \frac{1}{K}\sum_k \mbox{PMI-Score}[k]$;\\
            PMI\_sum $+=$ PMI\_score;\\
        }
        PMI\_final $\gets$ PMI\_sum / $iter$;\\
        \uIf{(PMI\_final $> maxPMI$)}{
            optVal $\gets$ topicNum; maxPMI $\gets$ PMI\_final;\\
        }
    }
    \textit{\#pragma omp parallel for reduction(max:topicNum})\\
    \For{$i \gets optVal- \left \lceil{span/2}\right \rceil$ \KwTo $optVal+\left \lceil{span/2}\right \rceil$}{
        Set $threadId \gets omp\_get\_thread\_num();$\\
        Set $tmp \gets threadId + optVal- \left \lceil{span/2}\right \rceil;$\\
        \uIf{($tmp == optVal$)}{
            break;\\
        }
        \uElse{
            repeat lines 4 \KwTo 6;\\
            Set $topicNum \gets tmp;$\\
            repeat lines 7 \KwTo 21;\\
        }
    }
    $optTopicNum \gets topicNum;$\\
    \end{algorithm}
    }
\setlength{\textfloatsep}{1pt}
    
	To estimate the most appropriate number of topics for our topic modeling approach, we propose a 2-step parallel algorithm. For the parallelization, we have employed OpenMP, a set of compiler directives and an API for our program (written in C++) that provides support for multi-platform, multiprocessing programming in shared-memory environments. OpenMP enabled us to write the algorithm so that the multithreading directives are skipped (or replaced with regular arguments) in the machines that do not have OpenMP installed. The designed algorithm to determine the optimal number for topics inference is provided in Algorithm \ref{alg:optTopNum}.

    The first step of our parallel algorithm takes an array of integers, $InputArr$. This array stores candidate number of topics, such as [$n_1$,$n_2$,...,$n_t$], where $n$ is an integer and $t$ is the \textit{array-size}. If $core$ refers to the number of CPU-cores available, it is advisable to limit $t$ within $[2, (\mbox{cores}-1)]$. This limit warrants that only one core would be assigned for each element in the array. Each core, in turn, builds $iter$ number of AOBTM models and calculates respective PMI-Scores. If we independently train AOBTM multiple times, on the same dataset with the same number of topics, we end up with slightly different PMI-Scores with each independent training. So, we designed our algorithm in such a way so that, for one candidate number of topics, each core builds $iter$ number of models and generates separate PMI-Scores. We stabilize the metric for corresponding candidate number by taking an average of the distinct PMI-Scores. We find the optimal candidate number ($optVal$) through \textit{reduction} (OpenMP operator) after all the threads finish their execution.
    
    In the second step of our algorithm, we finetune near the optimal candidate number ($optVal$) to determine the final value of the optimal topic number, $optTopicNum$. The user may specify $span$ with an integer, which defines the breadth of grid-search around $optVal$, observed in the algorithm's first step. Without any user specification, $span$ is automatically set as $core-1$. For each integer in the range of $span$ around $optVal$, we repeat the procedure of the first step to determine the optimal value and set it as the appropriate number of topics.
    
\begin{figure}[htbp]
        \centering
        \includegraphics[width=\linewidth]{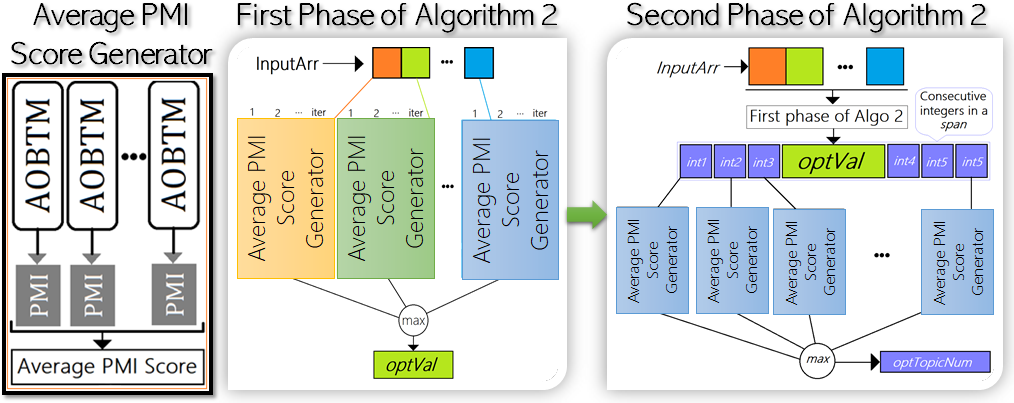}
\vspace{-2mm}
        \caption{Two Phases of Algorithm \ref{alg:optTopNum}, , determining the optimal topic number. Each element of the \textit{InputArr} is handled by one CPU-core. Circle-max represent the \textit{reduction} operator.}
        \label{fig:algo}
\vspace{-3mm}
    \end{figure}

    Fig. \ref{fig:algo} illustrate the first and second phase of Algorithm \ref{alg:optTopNum}, respectively. In essence, the first phase determines the appropriate number for topics inference ($optVal$) by evaluating the elements of the $InputArr$; the second phase determines the optimal topic number by evaluating integers around the $optVal$. The algorithm ensures that in each phase, one core evaluates only one integer topic number. 

\vspace{-2mm}
\subsection{Algorithm Determining Number of Versions to Consider}
\vspace{-1mm}
    Earlier, we have discussed how previous versions or time slices incur different contributions to the topic distributions of the latest time-slice. In AOBTM, we form the prior topic distributions for $t$-th time-slice by taking weighted contributions of the previous $win$ number of time slices into account. Users can define the parameter $win \in [1, v-1]$ in Algorithm \ref{alg:aobtm}, where $v$ denotes the available time slices. To make an educated decision about the parameter $win$, we can analyze the change in PMI Scores for different values of $win$. We build ($v-1$) number of AOBTM models for the latest time-slice with different values of $win$ and calculate the PMI-Scores. In the parallel block, we use OpenMp \textit{reduction} clause to find the maximized PMI-score. If OpenMp is not enabled in a machine, we store the scores in an array, where the score's index corresponds to the number of considered versions. Then, with one scan of the array, we determine the cutoff point where the PMI-score dropped and did not rise again. 
    
    We propose Algorithm \ref{alg:optWinNum} to determine the appropriate number of previous versions to consider automatically. 
    
    \scalebox{0.8}{
    \removelatexerror
    \begin{algorithm}[H]
    \caption{Optimal Number of Versions}
    \label{alg:optWinNum}
    \SetAlgoLined
    \SetKwInOut{Input}{Input}
    \SetKwInOut{Output}{Output}
    \SetKw{KwBy}{by}
    \SetKw{KwTo}{to}
    \Input{$dataset, v, iter$}
    \Output{$optVerNum$}
    Set $maxPMI \gets 0.0$, $optVerNum \gets 1;$\\
    \textit{\#pragma omp parallel for reduction(max:maxPMI})\\
    \For{$i \gets 1$ \KwTo $v-1$}{%
        Set $threadId \gets omp\_get\_thread\_num();$\\
        Set $win \gets threadId+1;$\\
        /* Each thread runs AOBTM $iter$ time by taking $win$
        number of previous versions into account,
        calculate the PMI-Scores and saves them in
        PMI\_final; (repeating lines 5 to 18 of Algo. 2);*/\\
        \uIf{(PMI\_final $> maxPMI$)}{
            optVerNum $\gets w$; maxPMI $\gets$ PMI\_final;\\
        }
    }
    \end{algorithm}
    }
\setlength{\textfloatsep}{0pt}
\vspace{-1.5mm}
\section{Experiments and Results}
\label{sec:experiments}
\vspace{-1mm}
    In this section, we evaluate the performance of AOBTM in identifying consistent and distinctive latent topics from corpora comprising of short text documents. We explain the datasets and compare the results of different topic modeling algorithms. Our focus is to answer the following research questions:

    \begin{itemize}
        \item \textbf{\textit{RQ1:}} Can AOBTM achieve better performance compared to other topic modeling methods?
        \item \textbf{\textit{RQ2:}} How do different parameter settings, document-lengths, and pre-processing approaches impact the performance of AOBTM?
        \item \textbf{\textit{RQ3:}} Using the parameters set by our parallel algorithms, how discriminative and coherent are the topics discovered by different topic modeling methods?
    \end{itemize}
\vspace{-2mm}
\subsection{Setups}
\vspace{-1mm}
\subsubsection{Datasets}
	To show the effectiveness of our approach, in addition to using app reviews, we use a large dataset of Twitter microblogs. Tweets are considered as short text and evaluation on this dataset can show the applicability of AOBTM on short text analysis.
	The details of the datasets are as follows: 
    	
    	\textit{App Reviews} from Apple Store and Google Play. We use the dataset provided by Gao et al. \cite{idea}, which is previously studied to evaluate AOLDA for extracting topics from app reviews. The dataset includes reviews that are related to a number of versions of the collected apps. 
    	The subject apps are distributed in different categories and platforms; this choice ensures the generalization of our approach. We enriched the provided datasets by adding the user-reviews collected from the latest versions of the subject apps.
    	However, one of the apps in this dataset, "Clean Master," was discontinued, and we could not acquire app-changelogs. Another app in the provided dataset, namely "eBay," had pulled enormous app-reviews from the app-stores. As the changelogs are critical to our evaluation metrics, we have decided to discard these two apps from the evaluation. We double-checked the provided app-reviews and changelogs in the dataset from the play stores and discarded the ones that could not be found.
    	Table \ref{table:sapps} summarizes the specifications of the app reviews datasets.
    	
    	\textit{Tweets2020} is a collection of approximately 200,000 tweets scraped from Twitter between January 1st and May 20th, 2020, where each month is considered as a time-slice. For the collection of the tweets, we have used an open-sourced twitter-scraper \footnote{https://pypi.org/project/twitter-scraper/}. We used 300 top trending topics over the region of North America to collect the tweets with timestamp. 
    	Besides the content, each tweet includes user id, timestamp, number of retweets, and likes.
    
	The user reviews and tweets collections contain many noisy words, such as repetitive words, casual words,  misspelled words, and non-informative words (e.g., "normally"). We have performed common text preprocessing techniques including removing meaningless words, lowercasing, lemmatization, digit and name replacement following  \cite{preprop}. 
	We apply the preprocessing technique in  for lemmatization and replace all digits with "\textless digit\textgreater." We also removed duplicate records and documents with a single word. 
    
    \begin{table}[htbp]
\vspace{-5mm}
    \caption{Subject apps from different app-stores}
\vspace{-4mm}
    \begin{center}
    \renewcommand{\arraystretch}{1.1}
    \resizebox{0.8\linewidth}{!}{%
    \begin{tabular}{lV{2.5}l|l|r|r}
        \hlineB{2.5}
        App Name & Category & Platform & \#Reviews & \#Versions\\
        \hlineB{2.5}
        {NOAA Radar} & {Weather} & {App Store} & {10,112} & {16} \\
        {Youtube} & {Multimedia} & {App Store} & {44,531} & {35} \\
        \hline
        {Viber} & {Communication} & {Google Play} & {19,327} & {9} \\
        {Swiftkey} & {Productivity} & {Google Play} & {23,121} & {17} \\
        \hlineB{2.5}
    \end{tabular}}
    \end{center}
    \label{table:sapps}
\vspace{-4mm}
    \end{table}
    
\subsubsection{Baselines} 
We select LDA \cite{lda}, BTM \cite{btm}, OLDA \cite{olda}, OBTM \cite{obtm}, and AOLDA \cite{idea} as our baseline methods to evaluate the performance of AOBTM. The details of the baseline algorithms are explained in Section \ref{sec:background}. 
All the experiments were carried on a Linux machine with Intel 2.21 GHz CPU and 16G memory. Following the literature \cite{idea}, we have used all algorithms implemented by Gibbs sampling in C++ \footnote{Code of BTM : http://code.google.com/p/btm/}.

\subsubsection{Evaluation Metrics}
    Good topic models deliver coherent \cite{tl2} and discriminative topics, which cover unique and comprehensive aspects of the corpus \cite{idea}. So, 
    We utilized PMI-Score as a measure of coherence \cite{btm} and Discreteness Score (Dis\_Score) to measure the discriminative property of the derived topics, which is inspired from the semantic similarity mapping in\cite{idea}. Higher values of PMI\_Score and Dis\_Score suggest the discovery of more coherent and discriminative topics. We also presented time-cost (seconds) per iteration (Time\_Cost in Table \ref{table:comp1}) as the third performance metric.
	We picked the top 10 terms from each generated topic to calculate the PMI-Scores, as explained in section \ref{sec:optTopNum}. For calculating Dis\_Score, we use Jensen Shannon (JS) Divergence $D_{JS}$ \cite{JSD}, to estimate the difference between two topic distributions ($\Phi$). The equations are provided below:
\vspace{-2mm}
    \begin{equation}
    \label{eq:diss}
\vspace{-2mm}
\resizebox{0.7\hsize}{!}{$
        \mbox{Dis\_Score} = \sum_{k=1}^K \bigg( \frac{\sum_{j=1,j\neq k}^K D_{JS}(\Phi_k^t||\Phi_j^t)}{K} \bigg)/K;
    $}
    \end{equation}
    \begin{equation}
    \label{eq:js}
\vspace{-2mm}
\resizebox{0.7\hsize}{!}{$
        D_{JS}(\phi_k^t || \phi_j^{t}) = \frac{1}{2}D_{KL}(\phi_k^t || M) + \frac{1}{2}D_{KL}(\phi_j^{t} || M);
    $}
    \end{equation}
    \begin{equation}
    \label{eq:kl}
\resizebox{0.7\hsize}{!}{$
        D_{KL}(P || Q) = \sum_i P(i)\log \frac{P(i)}{Q(i)};\;M = \frac{1}{2}(\phi_k^t + \phi_j^{t});
    $}
    \end{equation}
    
    Eq. \ref{eq:js} elaborates $D_{JS}$ of Eq. \ref{eq:diss}. Eq. \ref{eq:kl} defines the $D_{KL}$ (Kullback-Leibler Divergence) and $M$ from Eq. \ref{eq:js}. In Eq. \ref{eq:diss}, the innter term, $\sum_{j=1,j\neq k}^K D_{JS}(\Phi_k^t||\Phi_j^t)/K$ measures the difference of a single topic distribution's average with the rest of the topic distributions. In Eq. \ref{eq:kl}, $P(i)$ (or $Q(i)$) is the $i$-th item in $P$ (or $Q$).
    
    To provide a better comparison, we adopted three more performance metrics used in \cite{idea} to evaluate the performance of AOLDA.
    These metrics are $\mathrm{Precision_E}$, $\mathrm{Recall_L}$, and $\mathrm{F_{hybrid}}$. 
    Further details about these metrics are discussed in \cite{idea}. For app-reviews and twitter data, we have taken app-changelogs and popular hashtags, respectively as our ground-truths. 
    Here, $\mathrm{Precision_E}$ measures the accuracy in detecting emerging topics in the latest time slice, t \cite{idea}. $\mathrm{Recall_L}$ evaluates whether our prioritized topics (including both emerging and non-emerging) 
    reflect the changes mentioned in the change-logs or hashtags.
    Higher $\mathrm{F_{hybrid}}$ suggests that the change-logs and hashtags are more explicitly covered by detected topics. The higher score in $\mathrm{F_{hybrid}}$ also signifies that the prioritized issues reflect more of the change-logs and hashtags contents \cite{idea}.

\vspace{-2mm}  
\subsection{Result of RQ1: Comparison Results with Different Methods}
    Table \ref{table:comp1} presents the evaluation results, where, $P_E, R_L, F_h$ refer to $\mathrm{Precision_E}$, $\mathrm{Recall_L}$, and $\mathrm{F_{hybrid}}$, respectively.
    
    \begin{table}[htbp]
\vspace{-5mm}
    \caption{Comparison result of different methods}
\vspace{-4mm}
    \begin{center}
    \renewcommand{\arraystretch}{1.2}
    \resizebox{0.8\linewidth}{!}{%
    \begin{tabular}{m{1.54cm}V{2.5}c|c|m{0.6cm}|m{0.6cm}V{2.5}m{0.6cm}|m{0.6cm}|m{0.6cm}}
        \hlineB{2.5}
        {App-Name (\#Avg. Texts)} & {Methods} & {PMI\_Scores} & {Dis\_ Scores} & {Time\_ Cost} & {$P_E$} & {$R_L$} & {$F_h$} \\
        \hlineB{2.5}
        \multirow{6}{1.7cm}{Tweets2020 \\ (\texttildelow 39,803)} & {LDA} & {2.03 $\pm$ 0.04} & {0.68} & {43.27} & {NA} & {NA} & {NA} \\\cline{2-8}
        & {BTM} & {2.04 $\pm$ 0.02} & {0.61} & {46.8} & {NA} & {NA} & {NA} \\\cline{2-8}
        & {OLDA} & {1.90 $\pm$ 0.03} & {0.79} & {\textbf{17.11}} & {0.581} & {0.612} & {0.592}  \\\cline{2-8}
        & {OBTM} & {1.92 $\pm$ 0.03} & {0.79} & {21.91} & {0.574} & {0.605} & {0.588}  \\\cline{2-8}
        & {AOLDA} & {2.09 $\pm$ 0.03} & {\textbf{0.89}} & {56.7} & {0.603} & {0.677} & {0.657}  \\\cline{2-8}
        & {\textbf{AOBTM}} & {\textbf{2.13 $\pm$ 0.04}} & {0.82} & {63.87} & {\textbf{0.608}} & {\textbf{0.684}} & {\textbf{0.662}}  \\
        \hlineB{2.5}
        \multirow{6}{1.7cm}{NOAA Radar \\ (\texttildelow 632)} & {LDA} & {1.34 $\pm$ 0.03} & {0.57} & {18.8} & {NA} & {NA} & {NA} \\\cline{2-8}
        & {BTM} & {1.36 $\pm$ 0.03} & {0.63} & {22.31} & {NA} & {NA} & {NA} \\\cline{2-8}
        & {OLDA} & {1.38 $\pm$ 0.03} & {0.68} & {\textbf{8.25}} & {0.461} & {0.519} & {0.47}  \\\cline{2-8}
        & {OBTM} & {1.41 $\pm$ 0.03} & {0.71} & {14.6} & {0.576} & {0.495} & {0.548}  \\\cline{2-8}
        & {AOLDA} & {1.45 $\pm$ 0.04} & {0.75} & {23.44} & {0.568} & {0.492} & {0.533}  \\\cline{2-8}
        & {\textbf{AOBTM}} & {\textbf{1.48 $\pm$ 0.04}} & {\textbf{0.78}} & {20.54} & {\textbf{0.578}} & {\textbf{0.508}} & {\textbf{0.556}}  \\
        \hlineB{2.5}
        \multirow{6}{1.7cm}{Youtube \\ (\texttildelow 1,272)} & {LDA} & {1.56 $\pm$ 0.04} & {0.64} & {20.74} & {NA} & {NA} & {NA} \\\cline{2-8}
        & {BTM} & {1.62 $\pm$ 0.04} & {0.73} & {25.36} & {NA} & {NA} & {NA} \\\cline{2-8}
        & {OLDA} & {1.53 $\pm$ 0.03} & {0.76} & {\textbf{11.51}} & {0.439} & {0.455} & {0.448}  \\\cline{2-8}
        & {OBTM} & {1.55 $\pm$ 0.04} & {0.74} & {15.33} & {0.483} & {0.464} & {0.463}  \\\cline{2-8}
        & {AOLDA} & {1.61 $\pm$ 0.03} & {0.8} & {30.92} & {0.598} & {0.474} & {0.529}  \\\cline{2-8}
        & {\textbf{AOBTM}} & {\textbf{1.66 $\pm$ 0.03}} & {\textbf{0.82}} & {31.08} & {\textbf{0.615}} & {\textbf{0.482}} & {\textbf{0.538}}  \\
        \hlineB{2.5}
        \multirow{6}{1.7cm}{Viber \\ (\texttildelow 2,147)} & {LDA} & {1.65 $\pm$ 0.03} & {0.68} & {29.06} & {NA} & {NA} & {NA} \\\cline{2-8}
        & {BTM} & {1.72 $\pm$ 0.02} & {0.75} & {33.4} & {NA} & {NA} & {NA} \\\cline{2-8}
        & {OLDA} & {1.58 $\pm$ 0.03} & {0.8} & {\textbf{14.47}} & {0.458} & {0.395} & {0.419}  \\\cline{2-8}
        & {OBTM} & {1.62 $\pm$ 0.02} & {0.82} & {20.32}  & {0.417} & {0.308} & {0.365} \\\cline{2-8}
        & {AOLDA} & {1.76 $\pm$ 0.04} & {0.81} & {34.64} & {0.465} & {0.409} & {0.428}  \\\cline{2-8}
        & {\textbf{AOBTM}} & {\textbf{1.89 $\pm$ 0.03}} & {\textbf{0.85}} & {37.55} & {\textbf{0.572}} & {\textbf{0.411}} & {\textbf{0.508}}  \\
        \hlineB{2.5}
        \multirow{6}{1.7cm}{Swiftkey \\ (\texttildelow 1,360)} & {LDA} & {1.53 $\pm$ 0.02} & {0.67} & {23.18} & {NA} & {NA} & {NA} \\\cline{2-8}
        & {BTM} & {1.61 $\pm$ 0.02} & {0.68} & {28.85} & {NA} & {NA} & {NA} \\\cline{2-8}
        & {OLDA} & {1.42 $\pm$ 0.04} & {0.74} & {\textbf{16.06}} & {0.209} & {0.551} & {0.291}  \\\cline{2-8}
        & {OBTM} & {1.49 $\pm$ 0.04} & {0.73} & {21.55} & {0.313} & {0.52} & {0.392}  \\\cline{2-8}
        & {AOLDA} & {1.67 $\pm$ 0.03} & {0.76} & {29.48} & {0.526} & {0.631} & {0.584}  \\\cline{2-8}
        & {\textbf{AOBTM}} & {\textbf{1.71 $\pm$ 0.02}} & {\textbf{0.83}} & {31.26} & {\textbf{0.542}} & {\textbf{0.658}} & {\textbf{0.591}}  \\
        \hlineB{2.5}
    \end{tabular}}
    \end{center}
    \label{table:comp1}
\vspace{-3mm}
    \end{table}

	During evaluation for RQ1, we set the parameters as $w = 3$ and $K = 10$ for the adaptive online algorithms for the sake of uniformity. We have initialized $\alpha = 0.05$ and $\beta = 0.01$ for LDA based methods as they have achieved the best performance with these values for short texts in \cite{btm}. We have set $\alpha = 50/K$ and $\beta = 0.01$ for BTM based algorithms \cite{btm}.
	
    From Table \ref{table:comp1}, we observe that AOBTM delivers the highest PMI-Scores with every dataset by alleviating the data sparsity problem and considering the varying contributions of different time-slices or versions. 
    So, the topics discovered by AOBTM are more coherent and comprehensible. For discriminative topic learning, AOBTM performs better than other methods, except for the Tweets2020 dataset. A large amount of short texts per time-slice in the Tweets2020 dataset helps AOLDA to learn better document level word correlations and infer more discriminative topics; still, AOLDA did not generate higher PMI-Score than AOBTM for Tweets2020. 
	From the result, it is apparent that AOBTM exceeds the benchmark methods including its online version, OBTM, as well as AOLDA. Although AOBTM marginally improved the performance of AOLDA, the performance improvement of AOLDA over OLDA is approximately the same as that of AOBTM over OBTM.
	AOBTM also generated the highest scores for every dataset for $\mathrm{Precision_E}$, $\mathrm{Recall_L}$, and $\mathrm{F_{hybrid}}$, which indicates that our topic model can select emerging topics more precisely. 

    We acknowledge that AOBTM is more time-expensive than all the other baselines, but the runtime is comparable to adaptive online methods when the dataset is small. From Table \ref{table:comp1}, we observe that the difference of runtime between AOLDA and AOBTM is trivial; AOBTM even outperforms AOLDA in runtime for \textit{NOAA Radar} dataset, which has the lowest number of average short texts per version.
    
\vspace{-2.5mm}
\subsection{Result of RQ2: Effect of Different Parameter Settings, Document Lengths and Preprocessing Approaches}
\begin{figure}[htbp]
        \centering
        \includegraphics[width=0.6\linewidth]{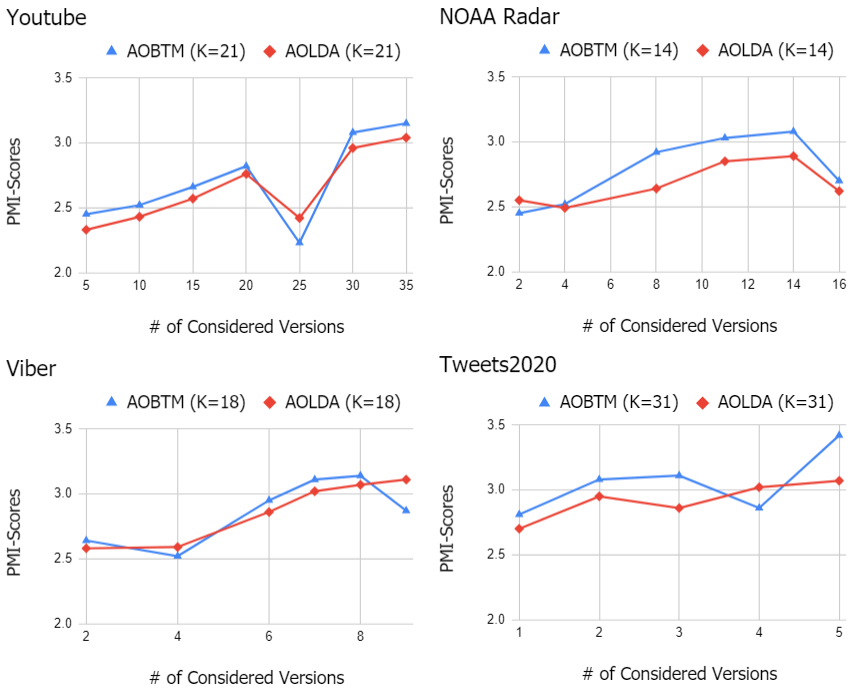}
\vspace{-2mm}
        \caption{PMI-Scores for varying number of considered versions or time-slices.}
        \label{fig:rq2w}
    \end{figure}

\begin{figure}[htbp]
        \centering
        \includegraphics[width=0.6\linewidth]{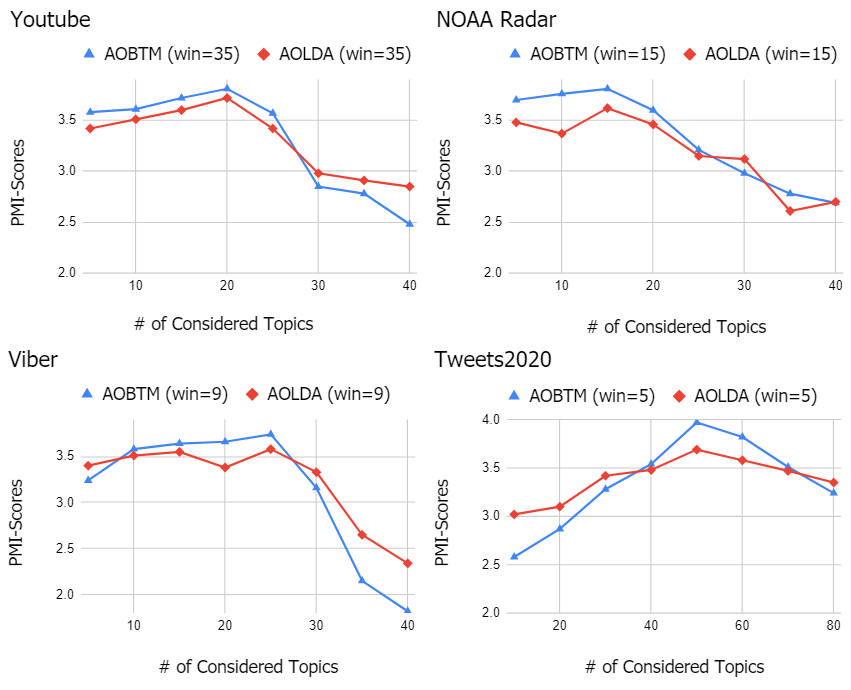}
\vspace{-2mm}
        \caption{PMI-Scores for varying number of topics.}
        \label{fig:rq2t}
    \end{figure}

\vspace{-1mm}
    \textbf{Effect of Different Parameter Settings. }
    In Fig. \ref{fig:rq2w} and Fig. \ref{fig:rq2t}, we have compared our method with AOLDA, as the number of previous versions to consider is unique to adaptive online algorithms. In Fig. \ref{fig:rq2w}, we consider distinct uniform topic-number for each dataset and calculate PMI-Scores for the different number of previous versions. The topic numbers for the dataset is calculated by Algorithm \ref{alg:optTopNum}. 
    In Fig. \ref{fig:rq2t}, we consider fixed window-size (number of versions) to calculate the PMI-Scores for varying number of topics. 
    
    We can observe that AOBTM in general generates the highest PMI-Scores, and the trendlines of both methods are analogous.
    In Fig. \ref{fig:rq2w}, the declines in the performance of AOBTM (i.e., win=25 in YouTube dataset) can transpire for the following reasons: the emergence of an unrelated novel topic in the recently considered versions (i.e., from 20th to 25th versions in YouTube dataset), content drifting, and higher occurrence of meaningless texts in the newly included versions \cite{obtm}. In Fig. \ref{fig:rq2t}, in all cases, the methods produced an increasing number of PMI-Scores until the tipping point generates the highest score. Once the methods reach their peak, a further increase in the number of topics generates incoherent coinciding topics inducing the reduction in PMI-Scores.
    
\vspace{-3mm}

\vspace{-2mm}
    
\begin{figure}[htbp]
        \centering
\vspace{-3mm}
        \includegraphics[width=0.5\linewidth]{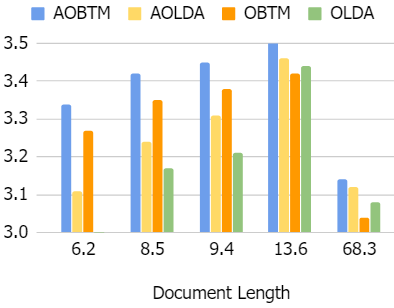}
        \caption{PMI-Scores for varying document lengths.}
        \label{fig:rq2d}
    \end{figure}

   \textbf{Effect of Document Length. }
    In Fig. \ref{fig:rq2d}, we have presented AOBTM's performance using PMI-Scores with respect to varying document lengths. We have considered average document length for each dataset to evaluate the considered methods. \textit{Tweets2020}, \textit{NOAA Radar}, \textit{Youtube}, \textit{Viber}, and \textit{Swiftkey} have average document length of 68.3, 8.5, 13.6, 9.4, and 6.2, respectively. AOBTM performs better than other methods for all datasets. It is worth noting that, as expected, LDA based methods performed well with large document length and bigger corpus, mostly because of the dataset's content richness and abundance of document level word co-occurrences.
    
    \textbf{Effect of Preprocessing. }
    In section \ref{sec:aobtm}, we have mentioned a preprocessing technique with \textit{Phrase Extraction} that can deliver more comprehensible top contributing terms in each discovered topic. 
    We implement AOBTM+ with the phrase extraction preprocessing technique. 
    In Table \ref{table:preprop}, we explicate how this technique in AOBTM+ generates better topic words than AOBTM with no phrase extraction. We selected a topic discovered from the Twitter dataset, which is related to \textit{Racism}. We can see that extracting phrases during preprocessing and training them with the rest of the terms help distinguishing key terms for topic representation, which is captured only by AOBTM+.
    
    \begin{table}[htbp]
\vspace{-4mm}
    \caption{Five most contributing terms from a topic from Tweets2020}
    \begin{center}
    \renewcommand{\arraystretch}{1.2}
    \resizebox{0.9\linewidth}{!}{%
    \begin{tabular}{c||l}
        \hline
        {Methods} & {Key-Terms} \\
        \hline
        {AOLDA} & {hate, race, black, white, stop}\\
        {AOBTM} & {black, hate, white, race, crime}\\
        {AOBTM+} & {stop\_racism, black, stop\_hate, police\_brutality, white\_supremacy}\\
        \hline
    \end{tabular}}
    \end{center}
    \label{table:preprop}
\vspace{-6mm}
    \end{table}

\subsection{Result of RQ3: Quality of Discovered Topics Using Parameters Determined by Proposed Algorithms}
    In \cite{numTop1, numTop2, numTop3, numTop4}, researchers have explored different ways to finetune the topic models' parameters. Their basic approach is to train various topic models (with different parameter settings) over several iterations to select one with the best performance. All the proposed procedures are computationally expensive, especially when executed sequentially. Moreover, all the existing libraries and packages that implement the mentioned procedures use LDA based models \cite{numTop5}. So, we proposed two parallel algorithms as described in Section \ref{sec:optTopNum} to determine 2 important parameters in our approach automatically: i) the number of topics to discover ($K$) and ii) the number of previous versions/time-slices to consider ($win$). In Table \ref{table:complexP}, time-complexities are provided for both algorithms. It is worth noting that the proposed algorithms run sequentially if the environment is not set up to perform parallelism.
    
    \begin{table}[htbp]
\vspace{-4mm}
    \caption{Time Complexities for Proposed Parallel Algorithms}
    \begin{center}
    \renewcommand{\arraystretch}{1.3}
    \resizebox{0.8\linewidth}{!}{%
    \begin{tabular}{cV{2.5}c}
        \hline
        {} & Time-Complexity\\
        \hlineB{2.5}
        {Algorithm \ref{alg:optTopNum}} & {$O\big([iter+span][N_{iter} K | N_B^{(t)}|+vKW]\big)$}\\
        {Algorithm \ref{alg:optWinNum}} & {$O\big([iter][N_{iter} K | N_B^{(t)}|+vKW]\big)$}\\
        \hline
    \end{tabular}}
    \end{center}
    \label{table:complexP}
\vspace{-4mm}
    \end{table}
    
\begin{figure}[htbp]
        \centering
        \includegraphics[width=0.6\linewidth]{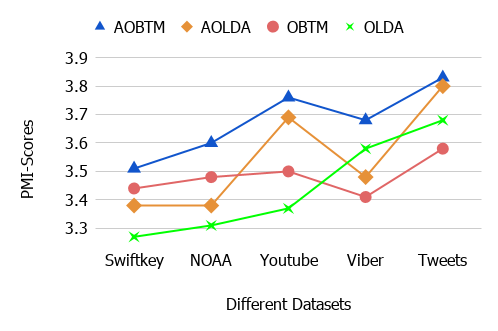}
        \vspace{-5mm}
        \caption{PMI-Scores generated by setting the parameters determined by Algorithms \ref{alg:optTopNum} and \ref{alg:optWinNum}}
        \label{fig:rq3}
\vspace{-1mm}
    \end{figure}

    We have employed the proposed algorithms to determine the best values for the parameters $K$ and $win$ for each dataset. For Tweets2020, Youtube, Viber, NOAA, and Swiftkey, Algorithm \ref{alg:optTopNum} determined the corresponding numbers of topics to be derived as 31, 22, 18, 13, and 11, respectively, whereas Algorithm \ref{alg:optWinNum} determined the best $win$ value to be considered as 5, 34, 9, 16, and 11, respectively.
    After setting the best detected parameters for the online and adaptively online algorithms, we have calculated the PMI-Scores for each dataset and present the results in Fig. \ref{fig:rq3}. The plot shows that AOBTM outperforms all the other online algorithms for all datasets. Compared to OBTM, AOLDA and OLDA perform better for datasets that contain longer documents. Furthermore, OBTM outperforms both of the LDA based methods when it comes to the limited datasets containing short texts.
\vspace{-1.5mm}
    
\subsection{Real-World Application}
\vspace{-1mm}
    Prompt detection of emerging topics from user reviews is crucial for app developers as these topics reveal users’ requirements, preferences, and complaints \cite{idea}.  Developers can proactively identify user-complaints and take quick actions to improve user experience through efficient analysis of the app-reviews. Timely and precisely identifying emerging issues helps developers to fix bugs, refine existing features, and add essential functions in the subsequent update of the application. 
    For this purpose, Gao et al. developed a framework named \textit{IDEA} to detect emerging issues from the app-reviews of popular applications \cite{idea}. \textit{IDEA} collects user reviews after the publication of different versions of the app and implement AOLDA to get the topic-word distributions for the app reviews collected after the publication of the latest version.
    We have modified the open-source framework IDEA and incorporated AOBTM instead of AOLDA to generate the topic-word distribution. The rest of the framework's components, such as preprocessing, emerging topic identification, and topic interpretation, remain the same. The modified version is denoted as OPRA (\textbf{O}nline A\textbf{p}p \textbf{R}eview \textbf{A}nalysis). 
\vspace{-4mm}   
    \begin{table}[htbp]
    \caption{Five most contributing terms from two sample topics}
    \begin{center}
    \renewcommand{\arraystretch}{1}
    \resizebox{0.5\linewidth}{!}{%
    \begin{tabular}{c||c|c}
        \hline
        Topics & IDEA & OPRA\\
        \hline
        \multirow{5}{10mm}{Topic 1} & {password} & {zoombomb}\\
        & {meeting} & {password}\\
        & {abuse} & {security}\\
        & {attack} & {policy}\\
        & {policy} & {disturb}\\
        \hline
        \multirow{5}{10mm}{Topic 2} & {message} & {group chat}\\
        & {status} & {message}\\
        & {channel} & {notification}\\
        & {chat} & {transfer}\\
        & {link} & {link}\\
        \hline
    \end{tabular}}
    \end{center}
    \label{table:ideap1}
\vspace{-5mm}
    \end{table}
    
    Inspired by the zoom case study as explained in Section \ref{sec;intro}, we have collected around 15,000 app reviews for \textit{Zoom Cloud Meetings} from Google Play. The average review length for this dataset is 7.8. These reviews were generated after the publication of the latest 5 versions of the app. For emerging topic detection, we set the parameters as $win = 3$ and $K = 10$ for fair evaluation. We also changed the initial values of $\alpha$ and $\beta$ to 0.1 and 0.01, respectively, as these values yielded best performance for IDEA (implementing AOLDA) in \cite{idea}.
    
    In Table \ref{table:ideap1}, we have reported top 5 most contributing words from two topics generated by IDEA and OPRA: first topic is closely related to app-security, and second topic is closely related to messaging feature of the app. In Table \ref{table:ideap2}, we have reported the corresponding PMI-Scores and Time-Cost for both frameworks. We can see that applying AOBTM in the framework slightly increases the time-cost, but generates more comprehensive and coherent topics.
    \begin{table}[htbp]
\vspace{-4mm}
    \caption{PMI-Scores \& Time Cost evaluation for Zoom app-reviews}
    \begin{center}
    \renewcommand{\arraystretch}{1.1}
    \resizebox{0.9\linewidth}{!}{%
    \begin{tabular}{c||c|c|c|c|c}
        \hline
        {Frameworks} & PMI Score & Time Cost & {$\mathrm{Precision_E}$} & {$\mathrm{Recall_L}$} & {$\mathrm{F_{hybrid}}$} \\
        \hline
        IDEA & 2.08 & 12.52 & 0.572 & 0.608 & 0.586\\
        OPRA & 2.35 & 16.8 & 0.593 & 0.619 & 0.608\\
        \hline
    \end{tabular}}
    \end{center}
    \label{table:ideap2}
\vspace{-6mm}
    \end{table}
    
\section{Related Work}
\label{sec:relatedWork}
    For the topic modeling for short texts, PLSA, LDA, and their variants suffer from the lack of enough word co-occurrences. To boost the performance of topic models, researchers had utilized external knowledge to produce supplementary essential word co-occurrences across short texts \cite{ttkalt, lcsst, ildamt}. The problem rests in that auxiliary information can be too scarce or too expensive (or both) for deployment.
    
	In the short text topic modeling regime,  Yin at al. introduced the DMM based topic modeling method in \cite{dmm}, where it is presumed that each short-text is sampled from only one latent topic. But this proved to be too simple and too strong of an assumption for any reasonable short text topic model \cite{cttm}.
    In self-aggregation based methods, short texts are merged into long pseudo-documents before topic inference to help develop rich word co-occurrence information. Researchers have used this type of method in \cite{satm, tmst}, where they have presumed that each short text is sampled from a long concatenated pseudo-document (unobserved in current text collection). This presumption enables inferring latent topics from long pseudo-documents. But the concatenation yielded suboptimal results in \cite{tmstswe, gfest} as merging short texts into long pseudo-documents using word embeddings cannot alleviate the loss of auxiliary information or metadata.    
    Global word co-occurrences based methods (i.e., \cite{btm, wntm}) try to use the rich global word co-occurrence patterns for inferring latent topics, where the adequacy of these co-occurrences alleviates the sparsity problem of short texts. \cite{btm} posits that the two words in a biterm share the same topic drawn from a mixture of topics over the whole corpus. This topic modeling algorithm is comparatively more robust and suitable for all the mentioned characteristics of short texts \cite{etmwe}.

    To solve the problems with streaming short texts and unordered topic generation, researchers proposed online models such as OLDA \cite{olda} and Online BTM \cite{btm}. In essence, online algorithms fit conventional topic models (i.e., BTM, LDA, respectively) over the data in a time-slice $t$ and use the inferred statistical data to adjust Dirichlet hyperparameters for the next time slice.

    Gao et al. \cite{idea} introduced the Adaptively Online LDA (AOLDA) by factoring in all the previous time-slices' contribution, instead of just the preceding one. Here, they have shown that the comparison among the topic distributions for more than two consecutive time-slices/versions can lead to more coherent and distinguishable topic learning. But this specific method uses LDA as their underlying topic model, which suffers from several discussed issues when it comes to short texts \cite{btm}.
    
\section{Threats to validity}
\label{sec:threats}
    \textbf{Human Evaluation.} In our experiments, we have only used deterministic scores to evaluate our results against the benchmarks. The authors of this paper have manually reviewed the outcomes of the considered topic models. We have consulted with fellow researchers about the model outcome and have considered the opinions of industry developers which confirmed that AOBTM generates more coherent key-words for extracted topics. But we did not perform any formal human evaluation to assess how well our model performs in practice compared to others.
    
    \textbf{Datasets.} 
    Our proposed topic models with text corpus distributed over different versions or time-slices. Evaluating our models using only version-tagged app-reviews from mobile applications that have multiple published versions in the app-store would not give us a precise idea about how this algorithm works with time slices. We have handled the problem by 
    We have evaluated our approach using a few mobile applications, which might affect the generality of our model. To migrate the problem, we have incorporated a new dataset with around 200,000 timeline-tagged tweets scraped by considering 300 top trending topics from Canada and the USA. Furthermore, we carefully selected apps so that we could demonstrate out topic model's performance for apps that have small or large number of reviews per version (\texttildelow 623-2,147) 
    
    \textbf{Ground Truth.} 
    To measure the extensibility of our topic model, we wanted to know how it scales to other online algorithms for prioritizing topics and detecting emerging ones. For selecting ground-truth, we have used key-terms from app-changelogs for app-reviews, as Gao et al. did in \cite{idea}. 
    In order to calculate precision, recall, and F-score, app-changelogs are used as ground truth, similar to \cite{idea}. However, we did not have any changelogs or tweet-summary to take as ground-truth for Tweets2020. We have manually selected top-trending hashtags over different time-slices to mine the key-terms. Other approaches for evaluation should be studied.
    
    \textbf{Memory and Time Cost.} We acknowledge that our model cannot compare to the benchmarks when it comes to memory and time-cost. Still, we are currently endeavoring to incorporate word-cooccurrence pattern algorithm to make our topic model faster while using significantly less resources.
        
\section{Conclusion and Future Work}
\label{sec:conclusion}
    
    In this paper, we proposed a novel adaptive topic modeling algorithm, AOBTM, which is able to discover coherent and discriminative topics from short texts. 
    AOBTM addresses the problems with conventional topic models by adopting a version sensitive strategy. 
    Along with AOBTM, we use a preprocessing technique that enables capturing distinguishable terms in an extracted topic. Moreover, we implemented two parallel algorithms to determine the value of the two most important parameters of our model automatically. The results of several experiments on app reviews and Twitter datasets confirm the performance of AOBTM compared to the state of the art algorithms. 
    
    We plan to improve the underlying BTM method using short text expansion and concept drifting detection and integrate it with a topic visualization tool specifically designed for app reviews. 
    For the parallel algorithms, we plan to use GPU-cores and shared memory cache to make the program run faster. We are currently endeavoring to incorporate word-cooccurrence pattern algorithm to make our topic model faster while using significantly less resources.

\newpage
\bibliographystyle{IEEEtran}
\bibliography{AOBTM}

\end{document}